\documentclass[AMS,STIX1COL]{WileyNJD-v2}
\usepackage{graphicx}
\usepackage{nicefrac}
\usepackage{amsmath}
\usepackage{amsfonts}
\usepackage{amsthm}
\usepackage{epsf}
\usepackage{bm}
\usepackage{longtable}
\AtBeginDocument{%

}
\newcommand{\g}{$g$~}

\newcommand{\exst}{${}^2 P_{3/2}~$}

\usepackage{braket}

\articletype{Conference Proceedings}%
\newcommand{\gd}{g_\mathrm{D}^{}}
\newcommand{\dgint}{\Delta g_{\mathrm{int}}}
\newcommand{\dgqed}{\Delta g_{\mathrm{QED}}^{}}
\newcommand{\dgrec}{\Delta g_{\mathrm{rec}}}
\newcommand{\dgns}{\Delta g_{\mathrm{NS}}^{}}
\newcommand{\dgintlo}{\dgint^{(0)}}
\newcommand{\dgintz}{\dgint^{(1)}}
\newcommand{\dgintzz}{\dgint^{(2)}}
\newcommand{\dgintzqed}{\dgint^{(1)}[\mathrm{QED}]}
\newcommand{\dgintps}{\dgint^{(1)}[+]}
\newcommand{\dgintns}{\dgint^{(1)}[-]}
\newcommand{\dgqedol}{\Delta g_{\mathrm{QED}}^{(1)}}
\newcommand{\dgqedtl}{\Delta g_{\mathrm{QED}}^{(2)}}
\newcommand{\dgse}{\Delta g_{\mathrm{SE}}^{}}

\newcommand{\dgvp}{\Delta g_{\mathrm{VP}}^{}}
\newcommand{\aZ}{\alpha Z}
\begin{document}

\title{\g factor of the $[(1s)^2(2s)^2 2p]~{}^2P_{3/2}$ state of middle-$Z$ boronlike ions}

\author[1,2]{V. A. Agababaev}
\author[1]{D. A. Glazov}
\author[1,3,4]{A. V. Volotka}
\author[1]{D. V. Zinenko}
\author[1]{V. M. Shabaev}
\author[5]{G. Plunien}

\address[1]{\orgdiv{Department of Physics}, \orgname{Saint-Petersburg State University}, \orgaddress{199034 Saint-Petersburg, \country{Russia}}}
\address[2]{\orgdiv{Saint-Petersburg State Electrotechnical University ``LETI''}, \orgaddress{197376 Saint-Petersburg, \country{Russia}}}
\address[3]{\orgdiv{Helmholtz-Institut Jena}, \orgaddress{D-07743 Jena, \country{Germany}}}
\address[4]{\orgdiv{GSI Helmholtzzentrum f\"ur Schwerionenforschung GmbH}, \orgaddress{D-64291 Darmstadt, \country{Germany}}}
\address[5]{\orgdiv{Instit\"ut f\"ur Theoretische Physik}, \orgname{Technische Universit\"at Dresden}, \orgaddress{D-01062 Dresden, \country{Germany}}}

\abstract[Abstract]{
Theoretical \emph{g}-factor calculations for the first excited \exst state of boronlike ions in the range $Z$=10--20 are presented and compared to the previously published values. The first-order interelectronic-interaction contribution is evaluated within the rigorous QED approach in the effective screening potential. The second-order contribution is considered within the Breit approximation. The QED and nuclear recoil corrections are also taken into account.
}
%
\maketitle
\section{Introduction}
Significant progress in the \emph{g}-factor studies in highly charged ions has been achieved in the last two decades \cite{sturm:17:a,shabaev:15:jpcrd}. Contemporary experiments have reached the precision of $10^{-9}$--$10^{-11}$ for hydrogenlike and lithiumlike ions \cite{haefner:00:prl,verdu:04:prl,sturm:11:prl,sturm:13:pra,wagner:13:prl}. One of the highlights in this field is the most accurate determination of the electron mass from the combined experimental and theoretical studies of the \g factor of hydrogenlike ions \cite{sturm:14:n}. Extension of these studies to lithiumlike ions has provided the stringent test of the many-electron QED effects \cite{wagner:13:prl,volotka:14:prl,koehler:16:nc,yerokhin:17:pra}. The high-precision $g$-factor measurement of the two isotopes of lithiumlike calcium \cite{koehler:16:nc} and the most elaborate evaluation of the nuclear recoil effect for this system \cite{shabaev:17:prl} have demonstrated a possibility to study the bound-state QED effects beyond the Furry picture in the strong field regime \cite{malyshev:17:jetpl}. It is expected that \emph{g}-factor studies in few-electron ions will be able to provide an independent determination of the fine structure constant $\alpha$ \cite{shabaev:06:prl,volotka:14:prl-np,yerokhin:16:prl}. 

The ALPHATRAP experiment at the Max-Planck-Institut f\"ur Kernphysik (MPIK) is capable of the ground-state \emph{g}-factor measurements for wide range of few-electron ions, including boronlike ones \cite{sturm:17:a}. The ARTEMIS project at GSI implements the laser-microwave double-resonance spectroscopy of the Zeeman splitting in both ground $[(1s)^2(2s)^2 2p]~{}^2P_{1/2}$ and first excited $[(1s)^2(2s)^2 2p]~{}^2P_{3/2}$ states of middle-$Z$ boronlike ions \cite{lindenfels:13:pra,vogel:18:ap}. In particular, boronlike argon is chosen as the first candidate for these measurements. Theoretical investigations of the \g factor of boronlike ions were performed recently in Refs.~\cite{glazov:13:ps,verdebout:14:adndt,shchepetnov:15:jpcs,marques:16:pra,agababaev:18:jpcs,maison:18:arxiv}. Various methods have been used in these works for evaluation of the interelectronic-interaction contribution, including the large-scale configuration-interaction approach in the basis of the Dirac-Fock-Sturm orbitals (CI-DFS) \cite{glazov:13:ps,shchepetnov:15:jpcs}, the GRASP2K \cite{verdebout:14:adndt} and MCDFGME \cite{marques:16:pra} packages based on relativistic multi-configuration Dirac-Hartree-Fock (MCDHF) method, the second-order perturbation theory (PT) in effective screening potential \cite{shchepetnov:15:jpcs,agababaev:18:jpcs}, and the high order coupled cluster (CC) method \cite{maison:18:arxiv}. For the ground-state \g factor of boronlike argon the results of the CI-DFS, PT, and CC approaches are in agreement on the level of $10^{-6}$, while both MCDHF results reveal a deviation on the level of $10^{-4}$. In the present work, we extend the second-order perturbation-theory calculations to the \exst state. The QED and nuclear recoil corrections are also taken into account. The results for boronlike ions in the range $Z$=10--20 are presented and compared to the previously published values~\cite{glazov:13:ps,verdebout:14:adndt,shchepetnov:15:jpcs,marques:16:pra,maison:18:arxiv}. We use the relativistic units ($\hbar = c = 1$) and the Heaviside charge unit ($\alpha=e^2/(4\pi)$, $e<0$) throughout the paper. 

\section{Methods and results}

The total \emph{g}-factor value of boronlike ion with zero nuclear spin can be written as
\begin{equation}
  g = \gd + \dgint + \dgqed + \dgrec + \dgns
\,,
\end{equation}
where $\dgint$, $\dgqed$, $\dgrec$, and $\dgns$ are the interelectronic-interaction, QED, nuclear recoil, and nuclear size corrections, respectively. The Dirac value $\gd$ for the $2p_{3/2}$ state is
\begin{equation}
  \gd = \frac{4}{15}\bigg[2\sqrt{4- (\alpha Z)^2}+1 \bigg] = \frac{4}{3} - \frac{2}{15}(\alpha Z)^2 - \dots
\,.
\end{equation}

The interelectronic-interaction correction is considered within the perturbation theory. The first-order term $\dgintz$ (one-photon exchange) is calculated within the rigorous QED approach, i.e., to all orders in $\aZ$. The second-order term $\dgintzz$ (two-photon exchange) is considered within the Breit approximation. The general formulae for this contribution can be found from the complete quantum electrodynamical formulae for the two-photon-exchange diagrams presented in Ref.~\cite{volotka:12:prl}. Care should be taken to account properly for the contribution of the negative-energy states, since it is comparable in magnitude to the positive-energy counter-part.

We incorporate the effective screening potential in the zeroth-order approximation. This improves the convergence of the perturbation theory and provides a reliable estimation of the higher-order remainder. The corresponding counter-terms should be considered in calculations of the first- and second-order contributions. The difference between the \emph{g}-factor values in the screening and pure Coulomb potentials is termed as the zeroth-order contribution $\dgintlo$. We use the following well-known screening potentials: core-Hartree (CH), Dirac-Hartree (DH), Kohn-Sham (KS), and Dirac-Slater (DS), see, e.g., Ref.~\cite{sapirstein:02:pra} for more details.  

In Table~\ref{tab:g-int} we present the interelectronic-interaction contributions to the \emph{g}-factor multiplied by $10^6$. The total value of $\dgint$ is found as,
\begin{equation}
  \dgint = \dgintlo + \dgintz + \dgintzz
\,,
\end{equation}
where the first-order correction $\dgintz$ is divided into the following three parts:
\begin{equation}
  \dgintz = \dgintps + \dgintns + \dgintzqed
\,.
\end{equation}
The positive-energy-states ($\dgintps$) and negative-energy-states ($\dgintns$) contributions are calculated in the Breit approximation. The QED contribution ($\dgintzqed$) is the difference between the rigorous QED and the Breit-approximation values. 

As the final results for $\dgint$, we take the values calculated in the Kohn-Sham potential. The uncertainty due to unknown higher-order contributions can be estimated as the spread of the obtained results for different potentials. As one can see from the Table~\ref{tab:g-int}, the maximal difference of the values of $\dgint$ varies between $1.8\times 10^{-6}$ for $Z$=10 and $0.8\times 10^{-6}$ for $Z$=20. Interelectronic-interaction corrections of the third and higher orders have been evaluated for lithiumlike ions within the CI-DFS \cite{volotka:14:prl} and CI \cite{yerokhin:17:pra} methods. The results obtained in these papers suggest that this estimation of the uncertainty is quite reliable.

The one-loop QED correction $\dgqedol$ is given by the sum of the self-energy and vacuum-polarization contributions,
\begin{equation}
  \dgqedol = \dgse + \dgvp
\,.
\end{equation}
The self-energy correction was calculated to all orders in $\aZ$ for both $2p_{1/2}$ and $2p_{3/2}$ states in the range $Z$=1--12 in Ref.~\cite{yerokhin:10:pra}. These values can be extrapolated to a good accuracy by the following $\aZ$-expansion \cite{yerokhin:10:pra,jentschura:10:pra},
\begin{equation}
\label{approx}
  \dgse = \frac{\alpha}{\pi} \left[ b_{00} + \frac{(\aZ)^2}{4}\,b_{20} + \frac{(\aZ)^4}{8}\,\left\{ \ln[(\aZ)^{-2}] b_{41} + b_{40} \right\} \right]
\,.
\end{equation}
The values $b_{00}(2p_{1/2})=-1/3$ and $b_{00}(2p_{3/2})=1/3$ have long been known~\cite{Brodsky:pr:1969,grotch:73:pra}. The values $b_{20}(2p_{1/2})=0.48429$ and $b_{20}(2p_{3/2})=0.59214$ have been found in Ref.~\cite{jentschura:10:pra}. Our fitting procedure based on the least squares method reproduces these coefficients on the level of $10^{-5}$ if they are taken as unknown, which serves as a check of its consistency. In this way we extrapolate the results of Ref.~\cite{yerokhin:10:pra} up to $Z$=20. In addition, we estimate the screening correction for the $2p_{3/2}$ state employing the effective nuclear charge $Z_\text{eff}$ instead of $Z$ in Eq.~(\ref{approx}). The effective nuclear charge $Z_\text{eff}$ is found from our rigorous calculations of the self-energy correction for the $2p_{1/2}$ state with an effective screening potential \cite{agababaev:18:jpcs}: Eq.~(\ref{approx}) with $Z_\text{eff}$ should reproduce the result obtained with the Kohn-Sham potential. The screening shift $Z-Z_\text{eff}$ lies in the range 1.3--1.7 for the ions under consideration. We ascribe the 100\% uncertainty to the screening correction obtained in this rather approximate way. 

The dominant contribution of the vacuum polarization is given by the two-electron diagrams where the vacuum-polarization potential acts on the $1s$ and $2s$ electrons. This contribution was estimated as $5.5\times 10^{-9}$ for $Z$=18 in Ref.~\cite{glazov:13:ps}, which is much smaller than the total theoretical uncertainty. The two-loop contribution $\dgqedtl$ is represented by its zeroth-order term of the $\aZ$-expansion~\cite{grotch:73:pra}.

The nuclear recoil effect in boronlike argon was calculated in Refs.~\cite{glazov:13:ps,shchepetnov:15:jpcs} within the Breit approximation to zeroth and first orders in $1/Z$. Systematic calculations of this effect for the $2p_{1/2}$ state in the range $Z$=10--20 were performed in Ref.~\cite{glazov:18:os}. Recently, these calculations have been extended to $Z$=20--92 including the leading-order QED contributions beyond the Breit approximation~\cite{aleksandrov:18:pra}. In the present paper, we evaluate this effect for the $2p_{3/2}$ state with the relativistic recoil operators to zeroth order in $1/Z$ with the Kohn-Sham effective screening potential.
The leading-order term of the finite-nuclear-size correction can be written as \cite{glazov:02:pla}
\begin{equation}
\label{dgns}
  \dgns = \frac{(\aZ)^6}{90} m^4 \langle r^4 \rangle
\,.
\end{equation}
For $Z$=10--20 it gives the values of the order $10^{-17}$--$10^{-15}$ which is negligible at the present level of accuracy.

The individual contributions and the total \emph{g}-factor values for the $2p_{3/2}$ state of boronlike ions in the range $Z$=10--20 are presented in Table~\ref{tab:g-int}. The values of $\dgint$ calculated in the Kohn-Sham potential are used. Our results for argon are in agreement with the PT results from Refs.~\cite{glazov:13:ps,shchepetnov:15:jpcs} and with the CC results from Ref.~\cite{maison:18:arxiv}. The difference between the data from Ref.~\cite{verdebout:14:adndt} and those of the present work ranges from $0.000\,042$ for $Z$=10 to $0.000\,094$ for $Z$=20. The difference between the data from Ref.~\cite{marques:16:pra} and those of the present work ranges from $0.000\,067$ for $Z$=14 to $0.000\,102$ for $Z$=20. The origin of this disagreement is not clear at present. We suppose that the negative-energy-states contribution was not taken into account completely in Refs.~\cite{verdebout:14:adndt,marques:16:pra}.  

Zeeman splitting of the $2p_{j}$ states acquires significant nonlinear contributions. In particular, the second- and third-order terms in magnetic field can be observed in forthcoming measurements for boronlike argon \cite{lindenfels:13:pra,glazov:13:ps}. Recently, the systematic calculations of these terms for the wide range of boronlike ions have been presented by our group~\cite{varentsova:18:pra}. The most important contribution for the $2p_{3/2}$ state is the shift of the levels with $m_j=\pm 1/2$ proportional to $B^2$. It can be represented as the $m_j$-dependent $g$-factor contribution varying from $\pm 1.52 \times 10^{-4}$ for $Z$=10 to $\pm 5.68\times10^{-6}$ for $Z$=20 at the field of 1\,T (it scales linearly with $B$). For more detailed description of the second- and third-order contributions see Ref.~\cite{varentsova:18:pra}.

\section{Conclusion}

In conclusion, the \emph{g} factor of the \exst state of boronlike ions in the range $Z$=10--20 has been evaluated with an uncertainty on the level of $10^{-6}$. The leading interelectronic-interaction correction has been calculated to all orders in $\aZ$. The higher-order interelectronic-interaction and nuclear-recoil effects have been taken into account within the Breit approximation. The one-loop self-energy correction has been found from extrapolation of the previously published high-precision results for $Z$=1--12 with an approximate account for screening.
%
%
\section*{Acknowledgements}

The work was supported in part by RFBR (Grants No.~16-02-00334 and 19-02-00974), by DFG (Grant No.~VO 1707/1-3), by SPbSU-DFG (Grant No.~11.65.41.2017 and No.~STO 346/5-1), and by SPbSU (COLLAB 2018: 34824940). V.A.A. acknowledges the support by the German-Russian Interdisciplinary Science Center (G-RISC). The numerical computations were performed at the St.~Petersburg State University Computing Center.

%
%
%
%

%
%
%
%
\clearpage
%
%
\LTcapwidth=\textwidth
\begin{longtable}{lrrrr}
\caption{\label{tab:g-int}
Interelectronic-interaction correction to the \emph{g} factor of boronlike ions in the \exst state. The terms of the zeroth ($\dgintlo$), first ($\dgintz$), and second ($\dgintzz$) orders of perturbation theory obtained with the core-Hartree (CH), Dirac-Hartree (DH), Kohn-Sham (KS), and Dirac-Slater (DS) screening potentials. First-order term is split into the contributions of the positive-energy ($\dgintps$) and negative-energy ($\dgintns$) spectra calculated within the Breit approximation and the QED part ($\dgintzqed$). All numbers are in units of $10^{-6}$.}
\\
\hline
& \multicolumn{1}{c}{CH} 
& \multicolumn{1}{c}{DH}
& \multicolumn{1}{c}{KS}
& \multicolumn{1}{c}{DS}
\\
\hline
\multicolumn{5}{c}{$Z = 10$}\\
\hline
$\dgintlo$&302.983&376.227&312.152&276.157\\
$\dgintps$ &  $-$22.611 &  $-$117.851  &$-$34.824 &  17.678 \\
$\dgintns$ & $-$22.371  & $-$2.806&  $-$19.623 & $-$29.741\\
$\dgintzqed$&  $-$0.141  & $-$0.111& $-$0.141 &  $-$0.158\\
$\dgintzz$&6.256&6.889&6.580&$-$0.758\\
$\dgint$&264.116&262.348&264.143&263.178\\
\hline
\multicolumn{5}{c}{$Z = 12$}\\
\hline
$\dgintlo$&368.346&462.044&379.402&334.176\\
$\dgintps$& $-$27.393& $-$148.063&  $-$41.786& 22.005\\
$\dgintns$& $-$29.402  &$-$4.746&$-$26.187& $-$38.614\\
$\dgintzqed$&$-$0.281&$-$0.233&$-$0.282&$-$0.308\\
$\dgintzz$&6.288&7.275&6.400&$-$0.441\\
$\dgint$&317.559&316.277&317.547&316.819\\
\hline
\multicolumn{5}{c}{$Z = 14$}\\
\hline
$\dgintlo$&433.852&547.842&446.756&392.322\\
$\dgintps$& $-$32.164& $-$178.105& $-$48.754& 26.389\\
$\dgintns$ & $-$36.370  &$-$6.680&$-$32.691& $-$47.407\\
$\dgintzqed$&$-$0.489&$-$0.420&$-$0.492&$-$0.528\\
$\dgintzz$&6.306&7.516&6.289&$-$0.265\\
$\dgint$&371.135&370.154&371.108&370.511\\
%
%
\hline
\multicolumn{5}{c}{$Z = 16$}\\
\hline
$\dgintlo$&499.514&633.710&514.250&450.613\\
$\dgintps$ & $-$36.888& $-$208.003& $-$55.691& 30.839\\
$\dgintns$ & $-$43.298  &$-$8.610 &$-$39.151&$-$56.142\\
$\dgintzqed$&$-$0.780&$-$0.686&$-$0.783&$-$0.833\\
$\dgintzz$&6.315&7.682&6.204&$-$0.166\\
$\dgint$&424.863 &424.092&424.828&424.311\\
\hline
\multicolumn{5}{c}{$Z = 18$}\\
\hline
$\dgintlo$&565.355&719.702&581.912&509.070\\
$\dgintps$& $-$41.544  &$-$237.766&$-$62.572 &35.370\\
$\dgintns$& $-$50.195  &$-$10.541&$-$45.575 &$-$64.829\\
$\dgintzqed$&$-$1.167&$-$1.043&$-$1.171&$-$1.235\\
$\dgintzz$&6.316&7.804&6.132&$-$0.116\\
$\dgint$&478.765&478.155&478.726&478.259\\
\hline
\multicolumn{5}{c}{$Z = 20$}\\
\hline
$\dgintlo$&631.397&805.864&532.824&567.713\\
$\dgintps$ & $-$46.116  &$-$267.395&$-$69.382 &39.996\\
$\dgintns$ & $-$57.065  &$-$12.472&$-$51.964 &$-$73.472\\
$\dgintzqed$&$-$1.661&$-$1.507&$-$1.668&$-$1.747\\
$\dgintzz$&6.311&7.896&6.065&$-$0.101\\
$\dgint$&532.866&532.386&532.824&532.134\\
\hline
\end{longtable}
%
%

%
\newpage
%
%
\begin{longtable}{lr@{}lr@{}l}
\caption{\label{tab:g}
Individual contributions to the \emph{g} factor of the \exst state of boronlike ions in the range $Z$=10--20. The values obtained with the Kohn-Sham potential are used for the interelectronic-interaction correction $\dgint$ (see Table~\ref{tab:g-int}). The \emph{g}-factor values from Refs.~\cite{verdebout:14:adndt,shchepetnov:15:jpcs,marques:16:pra,maison:18:arxiv} are given for comparison.}\\
%
%
\hline\\[-6pt]
& \multicolumn{2}{c}{${}^{20}_{10}$Ne$^{5+}$}
& \multicolumn{2}{c}{${}^{24}_{12}$Mg$^{7+}$}
\\[6pt]
\hline
Dirac value $\gd$                     &    1.&332\,623\,079     &    1.&332\,310\,417     \\
Interelectronic interaction $\dgint$  &    0.&000\,264\,1\,(18) &    0.&000\,317\,5\,(13) \\
One-loop QED $\dgqedol$               &    0.&000\,775\,7\,(5)  &    0.&000\,776\,3\,(7)  \\
Two-loop QED $\dgqedtl$               & $-$0.&000\,001\,2       &
$-$0.&000\,001\,2       \\
Nuclear recoil $\dgrec$               & $-$0.&000\,008\,9\,(15) & $-$0.&000\,007\,8\,(11) \\
\hline
Total value $g$                       &    1.&333\,652\,8\,(23) &    1.&333\,394\,1\,(17) \\
\hline
$g$ from Ref.~\cite{verdebout:14:adndt}
                                      &    1.&333\,695          &    1.&333\,448          \\[6pt]
\hline\\[-6pt]
& \multicolumn{2}{c}{${}^{28}_{14}$Si$^{9+}$} 
& \multicolumn{2}{c}{${}^{32}_{16}$S$^{11+}$}
\\[6pt]
\hline
Dirac value $\gd$                     &    1.&331\,940\,789     &    1.&331\,514\,136     \\
Interelectronic interaction $\dgint$  &    0.&000\,371\,1\,(10) &    0.&000\,424\,8\,(8)  \\
One-loop QED $\dgqedol$               &    0.&000\,777\,2\,(9)  &    0.&000\,778\,2\,(10) \\
Two-loop QED $\dgqedtl$               & $-$0.&000\,001\,2       & $-$0.&000\,001\,2       \\
Nuclear recoil $\dgrec$               & $-$0.&000\,006\,8\,(8)  & $-$0.&000\,006\,1\,(6)  \\
\hline
Total value $g$                       &    1.&333\,081\,1\,(16) &    1.&332\,709\,8\,(14) \\
\hline
$g$ from Ref.~\cite{verdebout:14:adndt}
                                      &    1.&333\,143          &    1.&332\,783          \\
$g$ from Ref.~\cite{marques:16:pra}
                                      &    1.&333\,148\,(7)     &    1.&332\,788\,(8)     \\[6pt]
\hline\\[-6pt]
& \multicolumn{2}{c}{${}^{40}_{18}$Ar$^{13+}$}
& \multicolumn{2}{c}{${}^{40}_{20}$Ca$^{15+}$}
\\[6pt]
\hline
Dirac value $\gd$                     &    1.&331\,030\,389     &    1.&330\,489\,471     \\
Interelectronic interaction $\dgint$  &    0.&000\,478\,7\,(6)  &    0.&000\,532\,8\,(7)  \\
One-loop QED $\dgqedol$               &    0.&000\,779\,5\,(12) &    0.&000\,780\,9\,(13) \\
Two-loop QED $\dgqedtl$               & $-$0.&000\,001\,2\,(1)  & $-$0.&000\,001\,2\,(1)  \\
Nuclear recoil $\dgrec$               & $-$0.&000\,004\,9\,(4)  & $-$0.&000\,004\,9\,(4)  \\
\hline
Total value $g$                       &    1.&332\,282\,5\,(14) &    1.&331\,797\,1\,(15) \\
\hline
$g$ from Ref.~\cite{verdebout:14:adndt}
                                      &    1.&332\,365          &    1.&331\,891          \\
$g$ from Ref.~\cite{marques:16:pra}
                                      &    1.&332\,372\,(1)     &    1.&331\,899\,(7)     \\
$g$ from Ref.~\cite{shchepetnov:15:jpcs}
                                      &    1.&332\,282\,(3)     &      &                  \\
$g$ from Ref.~\cite{maison:18:arxiv}
                                      &    1.&332\,286          &      &                  \\
\hline
\end{longtable}
\end{document}